\documentclass[letterpaper]{article}

\usepackage[T1]{fontenc}

\usepackage{geometry}
\usepackage{setspace}

\usepackage[style = chem-acs, doi = true, articletitle=true,]{biblatex} 
\usepackage{graphicx}
\usepackage{float}
\newfloat{scheme}{htbp}{los}
\floatname{scheme}{Scheme}
\floatname{chart}{Chart}
\newfloat{graph}{htbp}{loh}

\usepackage{chemformula} 
\usepackage[version = 4]{mhchem} 

\usepackage[hidelinks]{hyperref}   
\usepackage{mathtools}
\usepackage{tabularx}
\usepackage{gensymb}
\usepackage{enumitem}

\newcommand{\suppl}[1]{\href{https://doi.org/10.48550/arXiv.2605.14673}{[#1]}} 
\newcommand{\corrcol}{black}

\newcommand{\vcot}{\ensuremath{ \text{\textbf{Co}}_{ \text{\textbf{30 kV}}}^{\boldsymbol{\uparrow}}}}
\newcommand{\vcof}{$ \text{\textbf{Co}}_{\text{\textbf{5 kV}}}^{{\boldsymbol{\uparrow}}}$}
\newcommand{\vcoanneal}{\ensuremath{ \text{\textbf{Co}}_{\text{\textbf{anneal, 30 kV}}}^{\boldsymbol{\uparrow}}}}
\newcommand{\vfet}{\ensuremath{ \text{\textbf{Fe}}_{\text{\textbf{30 kV}}}^{\boldsymbol{\uparrow}}}}
\newcommand{\vfef}{\ensuremath{ \text{\textbf{Fe}}_{\text{\textbf{5 kV}}}^{\boldsymbol{\uparrow}}}}

\newcommand{\acot}{\ensuremath{ \text{\textbf{Co}}_{\text{\textbf{30 kV}}}^{\boldsymbol{\nearrow}}}}
\newcommand{\acof}{\ensuremath{ \text{\textbf{Co}}_{\text{\textbf{5 kV}}}^{\boldsymbol{\nearrow}}}}
\newcommand{\afet}{\ensuremath{ \text{\textbf{Fe}}_{\text{\textbf{30 kV}}}^{\boldsymbol{\nearrow}}}}
\newcommand{\afef}{\ensuremath{ \text{\textbf{Fe}}_{\text{\textbf{5 kV}}}^{\boldsymbol{\nearrow}}}}

\usepackage{authblk}

\author[1,$\ast$]{Aurys \v{S}ilinga}
\author[1]{Keir L. Edgar}
\author[2]{Andr\'as Kov\'acs}
\author[1]{Stephen McVitie}
\author[2]{Rafal E. Dunin-Borkowski}
\author[1]{Kayla Fallon}
\author[1,$\ast$]{Trevor P. Almeida}

\affil[1]{SUPA, School of Physics and Astronomy, University of Glasgow, Glasgow, G12 8QQ, United Kingdom}
\affil[2]{Ernst Ruska-Centre for Microscopy and Spectroscopy with Electrons, Forschungszentrum J\"ulich, 52425 J\"ulich, Germany}

\title{\textcolor{\corrcol}{Focused Electron Beam-Induced Deposition of Oblique Co and Fe Nanowires: Compositional and Magnetic Characterisation for Functional 3D Nanostructures}}

\date{*Email: a.silinga.1@research.gla.ac.uk, Trevor.Almeida@glasgow.ac.uk}

\begin{document}

\maketitle

\begin{abstract}
Focused electron beam induced deposition (FEBID) is an additive manufacturing technique uniquely suited for fabricating \textcolor{\corrcol}{functional 3D nanostructures} for a range of applications, including spintronic devices. However, the variation of growth dynamics associated with \textcolor{\corrcol}{lateral movement of the} electron beam results in structures with non-uniform composition when fabricating intricate 3D geometries. Herein, we measure changes in atomic composition and corresponding changes in magnetic induction in 3D ferromagnetic nanostructures with overhanging elements, e.g. bridges or arches. To investigate the effects of electron beam \textcolor{\corrcol}{lateral movement}, we fabricated Co and Fe nanowire (NW) structures with growth angle relative to the optic axis varying from 0° to 90°. The (scanning) transmission electron microscopy techniques of electron energy loss spectroscopy and off-axis electron holography were performed to map the NW elemental composition and magnetic induction as a function of NW growth angle. Comparison of the results reveals a reduction in metal content with increased growth angle in FEBID NWs. The magnitude of metal content reduction can be tuned by controlling electron beam parameters, and ferromagnetic NWs with approximately equal metal content at growth angles from 0° to 60° were fabricated by using the lowest viable electron beam voltage and the highest viable beam current to reduce the interaction volume and increase the metal content, respectively. \\
\includegraphics[width=8.25 cm]{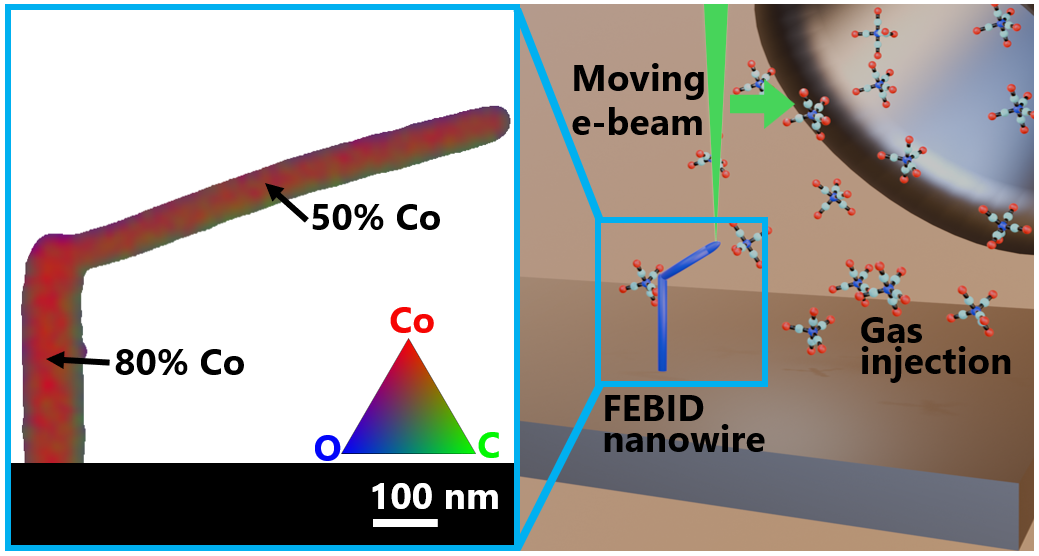}

\end{abstract}

\section*{Keywords}
nanostructure, ferromagnetism, FEBID, electron microscopy, electron holography, 3D, fabrication



\section{Introduction}
The environmental impact of conventional electronics can be mitigated by the development of spintronic computing architectures, which are theoretically more energy-efficient when handling digital data \cite{nikonov_power_2006}. Devices including magnetic racetrack memory for data storage \cite{parkin_magnetic_2008} or magnetic synapses for artificial neural networks \cite{ellis_machine_2023}  function by moving and storing data as magnetic domain walls in ferromagnetic nanowires (NWs). These designs have been tested in two-dimensional (2D) racetracks \cite{parkin_magnetic_2008}, but by expanding to three-dimensional (3D) architectures, the component density per chip area can be increased by angling NWs vertically with respect to the substrate \cite{gu_three-dimensional_2022}. Additionally, the shape anisotropy and symmetry breaking induced by NW curvature \cite{streubel_magnetism_2016, hertel_curvature-induced_2013} can be used to tune magnetic interactions that control magnetic domain wall (DW) movement. Complex 3D NW structures can be fabricated using focused electron beam induced deposition (FEBID), which is an additive manufacturing technique that can locally deposit ferromagnetic NWs with diameters down to 50 nm \cite{fernandez-pacheco_writing_2020, de_teresa_review_2016}. Even though the range of fabrication geometries is generally as unlimited as 3D-printing \cite{skoric_layer-by-layer_2020}, many functional ferromagnetic structures comprise NWs that are deposited at a constant angle to the substrate \cite{fernandez-pacheco_writing_2020, keller_direct-write_2018}. Since the purity and magnetic properties of the FEBID material are generally dependent on the orientation of the electron beam \cite{reisecker_review_2024}, movement of the beam can cause composition variations in the deposited nanostructures \cite{skoric_layer-by-layer_2020}. Ensuring that functional FEBID nanostructures have a uniform composition when fabricating complex geometries is crucial for the design of magnetic devices, therefore, deposition parameters are tested.

\subsection{FEBID chemical reactions}
FEBID is performed by injecting a precursor gas into the sample chamber of a scanning electron microscope (SEM) and irradiating solid material with the electron beam \cite{fernandez-pacheco_writing_2020, reisecker_review_2024}. This process drives the dissociation of precursor gas molecules primarily through irradiation by secondary electrons (SEs) emitted from the substrate, \textcolor{\corrcol}{and other forms of radiation present in the vicinity of the beam.} The reaction products \textcolor{\corrcol}{deposit on} surfaces in the vicinity of the reaction volume, typically forming a deposit of nanocrystalline \textcolor{\corrcol}{or amorphous} metal. The primary challenge associated with FEBID is fabricating nanostructures with desired geometries while maintaining high metal purity, e.g., Co, Fe, etc. Inclusion of C \& O contaminants and incomplete dissociation of the precursor gas is inherent to the FEBID process, which can result in relatively low metallic content \cite{utke_coordination_2022} and loss of ferromagnetic properties. Optimising the deposition parameters and inclusion of post-processing stages have been thoroughly investigated to produce deposits with predictable ferromagnetic and charge transport behaviours, required for reliable device operation \cite{cordoba_high-purity_2016, botman_creating_2009, magen_focused-electron-beam_2021}. The composition of FEBID material is attributed to the balance between multiple chemical reactions occurring at the site where the electron beam is incident on a substrate \cite{utke_coordination_2022}. This includes adsorption, which involves the formation of a bond between the precursor gas and the substrate; dissociation of the precursor molecules via bond cleavage; and the desorption of reaction products from the surface. The underlying chemical reactions have been studied both in isolation and as a combined deposition reaction. Co and Fe deposition is discussed in this work.

For the commonly used Co$_2$(CO)$_8$ precursor gas, autocatalytic adsorption has been observed to occur spontaneously \cite{muthukumar_spontaneous_2012}, but can also be driven by electron impact (EI) ionisation \cite{winters_ions_1965, utke_coordination_2022}. Regarding dissociation, mass spectroscopy experiments have shown that EI most frequently breaks bonds between Co and organic ligands, resulting in the creation of ions of the form Co$_2$(CO)$_8$ $\rightarrow$ Co$_{1-2}$(CO)$_{0-8}^{\pm}$ \cite{winters_ions_1965}. Calorimetry experiments have also shown that thermal dissociation (TD) at temperatures above 80~°C results in the deposition of pure Co and the desorption of CO gas, Co$_2$(CO)$_8$ $\xrightarrow{80\degree}$ 2Co$\downarrow$ $+$ 8CO$\uparrow$ \cite{connor_high_1973}. Although the FEBID material can contain organic ligands and is not guaranteed to be ferromagnetic, the conventional \textcolor{\corrcol}{beam-matter interactions} of ionisation by scattered electrons and beam-induced heating act to increase the metal content, which can be tuned by controlling the electron beam \cite{cordoba_high-purity_2016}.

For comparison, the Fe$_2$(CO)$_9$ precursor gas also undergoes EI ionisation of the form Fe$_2$(CO)$_9$ $\rightarrow$ Fe$_{1-2}$(CO)$_{0-9}^{\pm}$ \cite{markin_energy-resolved_2000}, and TD Fe$_2$(CO)$_9$ $\xrightarrow{280\degree}$ 2Fe$\downarrow$ $+$ 9CO$\uparrow$ above 280°C \cite{connor_high_1973}. Additionally, when Fe$_2$(CO)$_9$ is exposed to white light, heat, or electron beams \cite{bertini_time-dependent_2006}, it partially converts to Fe(CO)$_5$, which adsorbs autocatalytically \cite{hochleitner_electron_2008}. In practice, the deposition process for Fe and Co precursors is similar, but the higher TD temperature for Fe$_2$(CO)$_9$ requires the use of a higher intensity electron beam to promote TD when depositing Fe NWs. 

Experimentally, when the electron beam in an SEM is stationary or rastering over a flat substrate, the elemental composition, magnetisation, electrical resistivity, and geometry of the deposited structure can be reproducibly controlled \cite{de_teresa_review_2016}. This is achieved by tuning several deposition parameters, including the SEM accelerating voltage, beam current, the concentration of precursor gas, and the dwell time at each beam position. Using these methods, more than 90\% metal content has been achieved regularly in planar films and straight pillars \cite{botman_creating_2009, pablo-navarro_tuning_2017, serrano-ramon_ultrasmall_2011}. In comparison, when fabricating geometrically complex nanostructures, the position of the electron beam can be controlled precisely, but beam movement alters the radiation dose absorbed per unit volume during deposition. Maintaining high irradiation required to break metal-carbon bonds and deposit high-purity ferromagnetic metal is not straightforward when the beam transmits through a 3D sample and induces \textcolor{\corrcol}{codeposition on multiple surfaces}. Therefore, a trade-off must be made between beam energy, which affects the interaction volume, and beam current, which affects beam intensity.

The metal content of deposited NWs can be further improved by post-growth treatments, which include thermal annealing, electron beam curing, reactive gas exposure, and combinations of each \cite{botman_creating_2009, begun_post-growth_2015, puydinger_dos_santos_annealing-based_2016}. Thermal annealing has been the most extensively investigated and successful post-processing treatment for increasing the relative metallic content in FEBID deposits. Annealing temperatures in the 300-600 °C range are observed to induce thermal dissociation of carbonaceous material, which can diffuse and desorb from the deposit surface \cite{almeida_effect_2020, reisecker_review_2024}. This has been studied with Co \cite{martinez-perez_nanosquid_2018, pablo-navarro_purified_2018} and Fe \cite{pablo-navarro_situ_2019, shimojo_effects_2006} precursors, resulting in enhanced purity and crystallinity of FEBID deposits for most tested nanostructure geometries. Overall, FEBID of vertical NWs and flat films has been extensively studied to tune material functional properties, but FEBID material composition in oblique NWs is geometry-dependent and should be characterised to inform fabrication of bespoke 3D nanostructures.

\subsection{FEBID 3D printing}
\begin{figure}[h]
  \centering
    \includegraphics[width=15cm]{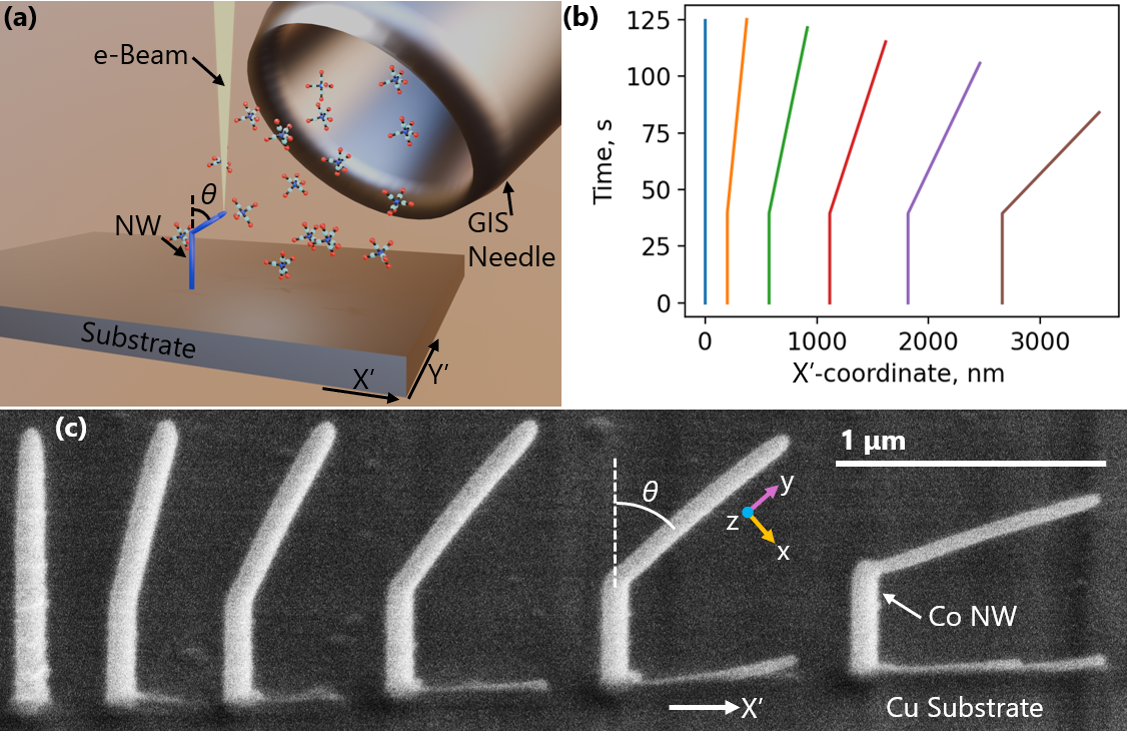}
  \caption{FEBID 3D printing process. (a) Schematic diagram of FEBID process, where NW structures are fabricated by injecting precursor gas through the GIS needle and \textcolor{\corrcol}{laterally moving} the electron beam along a path defined by stream files in SEM. (b) Streams that define beam position as a function of time, during deposition. (c) Tilt-corrected SE SEM image of FEBID Co NWs fabricated by executing the streams in (b). The inset shows the sample-aligned coordinate system used later in the analysis. To fabricate oblique nanostructures (c), the beam must be stationary at each position for an amount of time corresponding to (b).}
    \label{fig_febid}
\end{figure}

The electron beam in an SEM can be programmed to \textcolor{\corrcol}{move laterally} along a predefined path, which results in NWs growing at an angle ($\theta$) relative to the beam-axis, as illustrated in Figure \ref{fig_febid}a. $\theta = 0\degree$ corresponds to a NW that is deposited around a stationary electron beam, and, in this work, is approximately perpendicular to the substrate surface (vertical). A \textcolor{\corrcol}{laterally moving} beam can be used to fabricate intricate 3D NW structures, such as double-helices, helical lattices, Moebius strips, or NW cubes that show curvilinear magnetic interactions \cite{skoric_layer-by-layer_2020, fullerton_design_2025, sanz-hernandez_artificial_2020}, and may inform the design of spintronic devices. To fabricate such 3D nanostructures, predictive models are used to generate stream files that control the \textcolor{\corrcol}{lateral movement} and dwell time of the incident electron beam. Models based on Langmuir-type approaches \cite{toth_continuum_2015, sanz-hernandez_fabrication_2017} or atomistic Monte-Carlo \cite{prosvetov_atomistic_2022, de_vera_multiscale_2020} have been developed to enable FEBID 3D printing from CAD designs and fabrication of micrometre-sized FEBID structures from Co or Fe. These model types assume that material is deposited in thin layers without affecting the layers underneath, which \textcolor{\corrcol}{does not account for TD occurring in all volume heated by the beam and does not predict either the anisotropic growth of oblique NWs \cite{skoric_layer-by-layer_2020} or auto-catalysed crystalline growth \cite{hochleitner_electron_2008}.} Such models are most accurate when TD is limited, and hence lower-purity material is deposited. Methods have been developed to simulate thermal conductance and FEBID TD reactions \cite{skoric_layer-by-layer_2020, prosvetov_atomistic_2023}, but they have been tested so far with Pt and Au. 

In practice, the material properties of FEBID NWs are tuned depending on the intended application. Such optimisations leverage the observation that NWs exhibit characteristic responses to changes in the deposition parameters \cite{cordoba_high-purity_2016}. Since FEBID precursor gases for Co and Fe are chemically similar and the underlying electron scattering interactions are the same \cite{randolph_focused_2006}, the effects of changing deposition parameters are also similar. They can be observed by depositing two NWs side-by-side; starting with NW 'A', changing one deposition parameter, and then depositing NW 'B'. The observed characteristic effects are:
\begin{enumerate}[label=E\arabic*., start=0]
    \item When nothing is changed, A and B are approximately identical. This indicates that the deposition is consistent, and is a necessary condition for the accuracy of FEBID CAD modelling \cite{toth_continuum_2015,skoric_layer-by-layer_2020, prosvetov_atomistic_2022}.
    
    \item When the beam current is increased, B will have a larger volume and higher metal content than A, as demonstrated for Co and Fe \cite{magen_focused-electron-beam_2021, pablo-navarro_tuning_2017, fernandez-pacheco_magnetotransport_2009}. There is an upper limit to the beam current beyond which randomly oriented crystalline growth is observed \cite{vollnhals_electron-beam_2014, utke_cross_2005, bertini_time-dependent_2006, hochleitner_electron_2008}. 
    
    \item When the beam is defocused, B will have a larger diameter and lower metal content than A. This has been observed for Co, Fe, and Pt \cite{utke_high-resolution_2002, randolph_focused_2006, cordoba_high-purity_2016, plank_influence_2008}

    \item When B is annealed at a temperature in the 300-600~\textdegree C range, its metal content will increase, as seen for both Co and Fe \cite{martinez-perez_nanosquid_2018, pablo-navarro_purified_2018, pablo-navarro_situ_2019, shimojo_effects_2006}.
    
    \item When the beam accelerating voltage is increased, the geometry of B will be different from A. If the SEM beam is \textcolor{\corrcol}{moving laterally} during deposition, the cross-section of B will be more elongated (elliptical) than the cross-section of A. This has been observed for Fe, Co, FeCo$_3$, and Pt \cite{winkler_shape_2021, keller_direct-write_2018, plank_influence_2008, hochleitner_electron_2008, skoric_layer-by-layer_2020}.

    \item If the \textcolor{\corrcol}{lateral movement} speed of the beam is increased, the metal content of B will decrease relative to A. This is measured for Co and Fe NWs in this work.
\end{enumerate}
In this work, Co NWs are fabricated with $\theta=0\degree$ to test effects E0-E3 and verify that the FEBID system is functioning normally. Then Co NWs and Fe NWs are fabricated with $0\degree \leq \theta <90\degree$ at different accelerating voltages to quantify effects E4 \& E5 and test whether E5 is material-agnostic. Furthermore, correlations among $\theta$, metal content, and magnetic induction in ferromagnetic NWs are measured; to develop methods for improving oblique NW composition uniformity and inform the development of FEBID 3D-printing algorithms for ferromagnetic nanostructures.
    
\section{Methods}

\subsection{Sample fabrication}
A dual-beam Thermo Fisher Scientific HELIOS Xe plasma focused ion beam (FIB) SEM was used to ion-mill flat substrates from a copper film, which could be electrically grounded during TEM and SEM imaging. NWs were deposited on the copper substrates using FEBID. The gas injection system (GIS) heated solid precursor material to induce sublimation, and the precursor gas was injected through the \textcolor{\corrcol}{GIS needles positioned at an approximately 250~µm height above the substrate surface and a 100~µm lateral distance from the deposition point. Co and Fe needles were inserted at 40° and 60° angles relative to the SEM beam, respectively. Needle orientation in the lateral plane was at a 50° angle clockwise from the +X' direction.} For FEBID, the Co$_2$(CO)$_8$ and Fe$_2$(CO)$_9$ precursors were heated to 29° C and 37° C, respectively. Nine sets of FEBID NWs were fabricated: first, five sets to characterise NWs deposited using a stationary electron beam as reference samples; and secondly, another four sets to measure the composition change in NWs deposited with a \textcolor{\corrcol}{laterally moving} electron beam. Table \ref{table1} lists the details of each sample set, including the precursor gas, beam accelerating voltage and current. For completeness, the base pressure in the SEM chamber was recorded before and during deposition, and to indicate the reaction rate, the vertical NW height to deposition time ratio (growth rate) is recorded. Compositional analyses and beam conditions for each individual NW are provided in Supporting Information \suppl{1}. The arrows in the dataset titles indicate whether the deposition was at $\theta=0\degree$~($\uparrow$) or $\theta>0\degree$~($\nearrow$). All NWs were deposited sequentially, from left to right. For each deposition, the electron beam was focused to achieve minimum NW diameter. The samples were plasma-cleaned with an oxygen-argon 1:3 mixture for 15~s to reduce amorphous carbon growth during scanning TEM (STEM) imaging. \textcolor{\corrcol}{Calibration EELS measurements showed that 15~s plasma-cleaning removed an approximately 5~nm thick layer from the carbon-rich Co NW surface and changed the number of O and Co atoms by less than 1\%.} 

\begin{figure}[H]
  \centering
    \includegraphics[width=15cm]{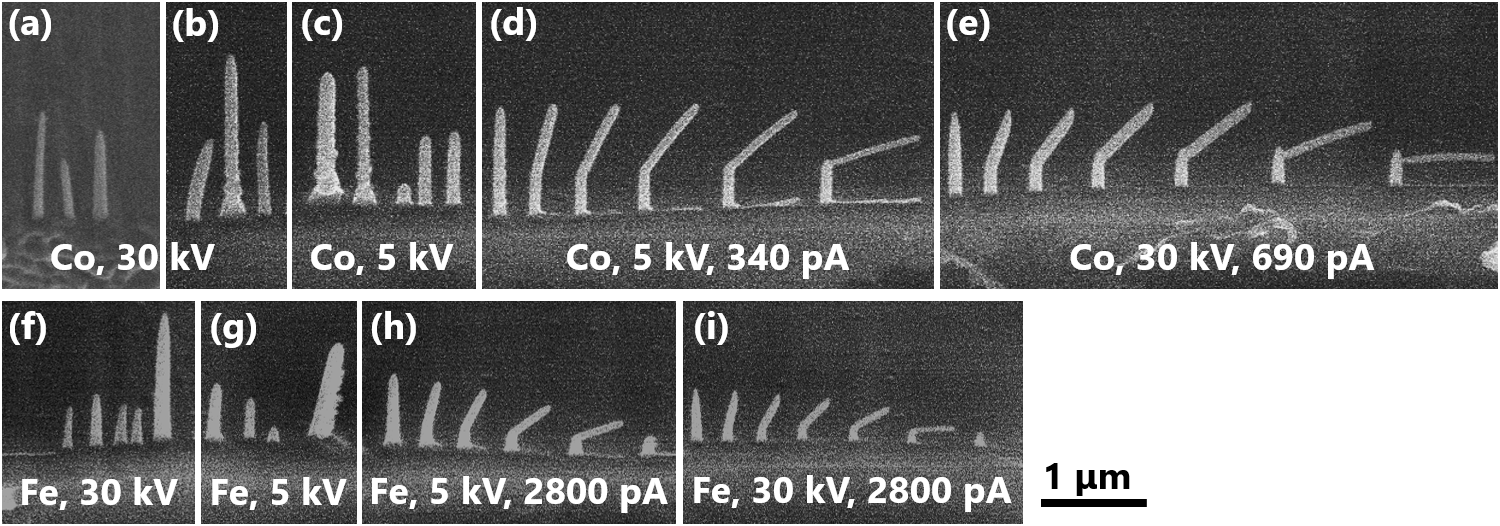}
  \caption{SE SEM images of FEBID NWs that are characterised and compared. (a) \vcoanneal. (b) \vcot. (c) \vcof. (d) \acof. (e) \acot. (f) \vfet. (g) \vfef, (h) \afet. (i) \afef. All images were acquired with the sample tilted 55$\degree$ and are shown at the same scale. High-resolution images \textcolor{\corrcol}{and deposition parameters for each NW} are available in the Supporting Information \suppl{1}.}
  \label{fig_samples}
\end{figure}

The NW structures were fabricated using stream files that encode the beam's X'- and Y'-deflections (in the substrate plane) as functions of deposition time. Vertical NWs were deposited using a stationary beam for 120~s, whilst oblique NWs were deposited by maintaining the beam stationary for 40~s and then \textcolor{\corrcol}{moving} it laterally in the X'-direction at a constant rate for up to 80~s. \textcolor{\corrcol}{The stream files were generated using the F3ast software package \cite{skoric_layer-by-layer_2020} and were calibrated to increase X'-deflection in approximately 2.5~nm steps at rates from 0 to 8 steps per second.} The electron beam dwell time at each coordinate is shown in Figure \ref{fig_febid}b. For the \acof$ $ dataset, each stream in Figure \ref{fig_febid}b moves the SEM beam at a different rate; therefore, the deposited NWs in Figure \ref{fig_febid}c grow at different angles relative to the vertical NW. Samples from all datasets are shown in Figure \ref{fig_samples}. Datasets \vcot, \vcof, \vcoanneal, \vfet, and \vfef are used to find deposition parameters that enable controlled deposition of vertical ferromagnetic NWs. Datasets \acof, \acot, \afef, and \afet$ $ use the parameters that deposit ferromagnetic material to fabricate oblique NW structures. For calibration, NWs in the dataset \vcoanneal$ $ were fabricated with lower currents and gas pressures, and characterised before and after annealing at 350~°C for 30~min to verify consistency with previous works \cite{puydinger_dos_santos_annealing-based_2016, pablo-navarro_purified_2018, magen_focused-electron-beam_2021, cordoba_high-purity_2016}. \textcolor{\corrcol}{The NWs were annealed in situ at high vacuum using a Gatan 628 single-tilt TEM heating holder.} In summary, stream files are defined to fabricate oblique FEBID NWs with growth angles ranging from 0° to 90°, and are executed using different precursor materials, accelerating voltages, and beam currents.

The precursor gas pressure was held constant during stream file execution, with fluctuations not exceeding $2 \cdot10^{-5}$~Pa, which \textcolor{\corrcol}{accounts for pressure changes of less than 15\% and is assumed} to have negligible effects on composition or growth rate \cite{pablo-navarro_tuning_2017}. The spatial distribution of precursor gas concentration depends on GIS alignment \cite{wanzenboeck_mapping_2014} and can affect NW deposition reaction rate, geometry, and atomic composition \cite{pablo-navarro_tuning_2017}. Although local gas concentrations were not measured, they are assumed to be constant because GIS alignment and chamber pressure were constant during deposition.

\begin{table}[H]
\caption{Deposition parameters for FEBID samples. Each dataset corresponds to a series of NWs deposited by executing the stream files in one area of the sample.\label{table1}}
\begin{center}
\begin{tabularx}{16.5 cm}{|p{3cm} | p{1.2cm}| X| X| X| X| X| X|}

 \hline
 Dataset title & Beam \space\space\space\space voltage (kV) & Beam \space\space\space\space\space\space\ current (pA) & Deposition pressure (Pa) & Base \space\space\space\space\space\space\space\space pressure (Pa) & Growth rate \space\space\space (nm/s) & Precursor gas \\ 
 \hline\hline
    $\boldsymbol{\uparrow} \text{\textbf{Co}}_{\text{\textbf{anneal, 30 kV}}}$ & $30$ & $21-340$ & $0.9 \cdot 10^{-4}$ & $0.5 \cdot 10^{-4}$ & $6-10$ & Co$_2$(CO)$_8$ \\ 
 \hline
  $\boldsymbol{\uparrow} \text{\textbf{Co}}_{\text{\textbf{30 kV}}}$ & $30$ & $170-2800$ & $2.2 \cdot 10^{-4}$ & $0.7 \cdot 10^{-4}$ & $8-26$ & Co$_2$(CO)$_8$ \\ 
 \hline
  $\boldsymbol{\uparrow} \text{\textbf{Co}}_{\text{\textbf{5 kV}}}$ & $5$ & $170-2800$ & $2.2 \cdot 10^{-4}$ & $0.7 \cdot 10^{-4}$ & $11-22$ & Co$_2$(CO)$_8$ \\ 
 \hline
   $\boldsymbol{\uparrow} \text{\textbf{Fe}}_{\text{\textbf{30 kV}}}$ & $30$ & $170-11000$ & $1.5 \cdot 10^{-4}$ & $0.2 \cdot 10^{-4}$ & $6-20$ & Fe$_2$(CO)$_9$ \\ 
 \hline
   $\boldsymbol{\uparrow} \text{\textbf{Fe}}_{\text{\textbf{5 kV}}}$ & $5$ & $170-11000$ & $1.2 \cdot 10^{-4}$ & $0.2 \cdot 10^{-4}$ & $6-20$ & Fe$_2$(CO)$_9$ \\ 
 \hline
  $\boldsymbol{\nearrow} \text{\textbf{Co}}_{\text{\textbf{30 kV}}}$ & $30$ & $690$ & $2.2 \cdot 10^{-4}$ & $0.7 \cdot 10^{-4}$ & $8$ & Co$_2$(CO)$_8$ \\ 
 \hline
  $\boldsymbol{\nearrow} \text{\textbf{Co}}_{\text{\textbf{5 kV}}}$ & $5$ & $340$ & $2.2 \cdot 10^{-4}$ & $0.7 \cdot 10^{-4}$ & $11$ & Co$_2$(CO)$_8$ \\ 
 \hline
   $\boldsymbol{\nearrow} \text{\textbf{Fe}}_{\text{\textbf{30 kV}}}$ & $30$ & $2800$ & $0.8 \cdot 10^{-4}$ & $0.2 \cdot 10^{-4}$ & $5$ & Fe$_2$(CO)$_9$ \\ 
 \hline
   $\boldsymbol{\nearrow} \text{\textbf{Fe}}_{\text{\textbf{5 kV}}}$ & $5$ & $2800$ & $1.0 \cdot 10^{-4}$ & $0.2 \cdot 10^{-4}$ & $7$ & Fe$_2$(CO)$_9$ \\ 
 \hline
\end{tabularx}
\end{center}
\end{table}

\subsection{Elemental mapping}
A JEOL ARM200cf operated at 200 kV (University of Glasgow) equipped with a cold field emission gun, a CEOS (Corrected Electron Optical System GmbH) probe corrector and Gatan 965 Quantum ER spectrometer was used to perform STEM electron energy loss spectroscopy (EELS) elemental mapping. A dual EELS system with a high-speed electrostatic deflector \cite{craven_fast_2002, craven_dual_2008} allowed simultaneous acquisition of high-loss and low-loss spectra from select areas of the FEBID NWs.  The spectra were acquired using a 0.5 eV/pixel dispersion, 2.5 mm aperture, 29~mrad convergence semi-angle and 36~mrad collection semi-angle. Principal component analysis (PCA) \cite{lucas_multivariate_2013} was utilised to remove X-ray signatures and reduce noise in high-loss spectra by reconstructing from the first 150 components, such that PCA would alter quantification results by less than 0.1\% in sample areas with high signal-to-noise ratio (SNR), but provide smoothing in areas of low SNR. Spectra with more than $10^4$ counts per energy channel in the fitting region were quantified using the Gatan Digital Micrograph elemental quantification plugin \cite{scott_near-simultaneous_2008}. \textcolor{\corrcol}{Supporting Information Figure \suppl{S10} compares high-loss EELS spectra from 20~nm wide centre bands of NWs with 130~nm and 30~nm projected thicknesses, showing that sufficient EELS signal was acquired for all NWs.} The measurement error stems from the combined contributions of Poisson noise and uncertainties of fitting the Hartree-Slater cross sections \cite{williams_high_2009}. The uncertainty for relative atomic content measurement (atom \%) in the thickest part of the NWs is less than 3\% per scanned pixel. 

\begin{figure}[h]
  \centering
    \includegraphics[width=15cm]{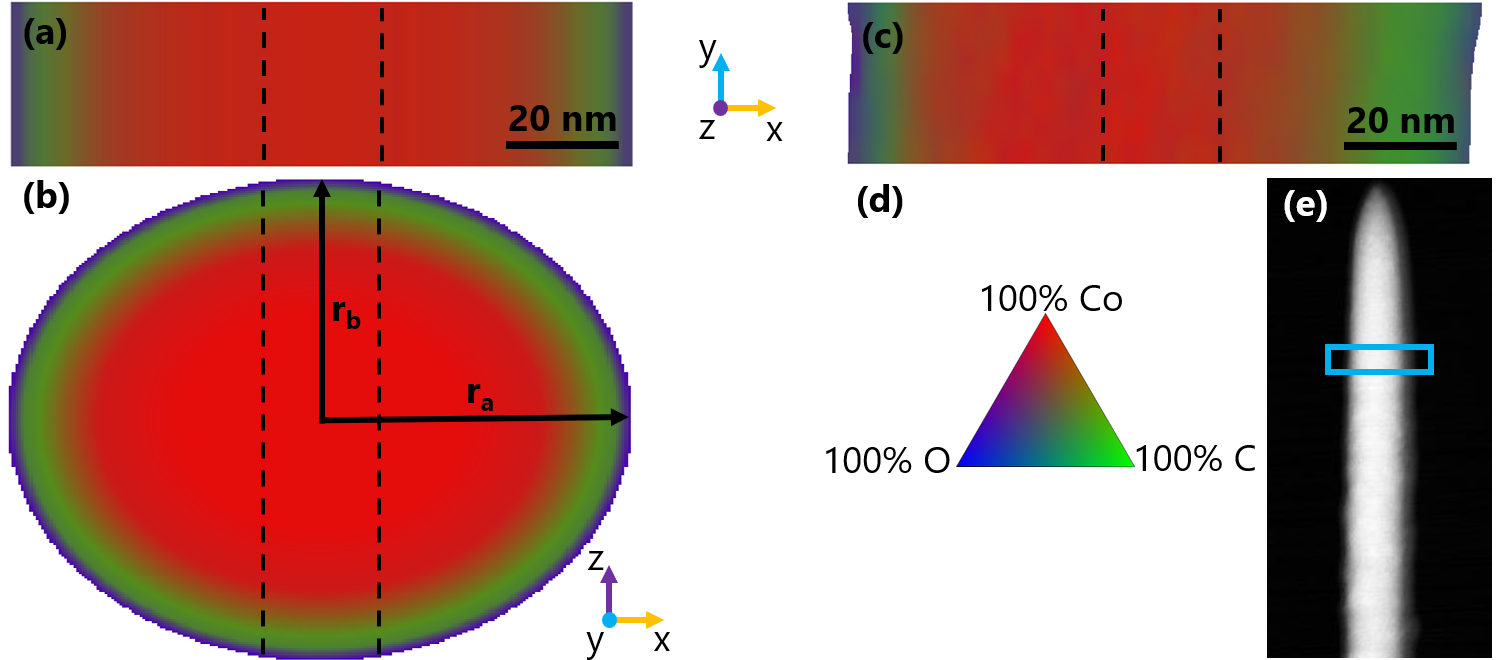}
  \caption{\textcolor{\corrcol}{Assumed} model of NW atomic composition. (a) Calculation of a compositional map, assuming the NW cross-section (b). (b) Model of an elliptical cross-section with a pure Co core, intermediate layers, and outer layers rich in C and O. (c) STEM EELS compositional map for a vertical NW in dataset \vcof. (d) Colour map for (a), (b), and (c). (e) STEM annular dark field (ADF) image identifying the location where (c) was acquired.  FEBID NWs are modelled as multi-layered structures with elliptical cross-sections. Black arrows indicate the semi-axes of the ellipse, the insets show the coordinate system, and the dashed lines identify the volume where composition was measured.}
  \label{fig_model}
\end{figure}

A Cartesian coordinate system was defined to describe compositional distributions in oblique cylindrical NWs (Figure~\ref{fig_febid}c inset), with the z-axis parallel to the TEM beam, the y-axis parallel to the NW length, and the x-axis parallel to the projected NW diameter. Figure \ref{fig_model} provides an example. The variation along the x-axis in the calculated EELS map Figure~\ref{fig_model}a corresponds to the projected distribution of atomic composition for a model NW with a radially varying composition and an elliptical cross-section, shown in Figure~\ref{fig_model}b. FEBID NWs exhibit a radially layered structure, consistent with a Co (or Fe) rich core and an outer shell rich in C and O. Hence, care must be taken when interpreting EELS maps, as they represent the thickness projection of a radially varying structure, as previously observed \cite{castillo-rico_stopping_2021}. This model is qualitatively consistent with experimental EELS maps of straight NWs with smooth surfaces. Such as the example in Figures \ref{fig_model}c and \ref{fig_model}e, deposited with a 340 pA beam current in the \vcof$ $ sample. \textcolor{\corrcol}{However, one EELS map acquired in projection is insufficient to reconstruct the 3D distribution of atomic composition, thus, the core-shell structure is assumed based on cross-sectional measurements in previous works \cite{magen_focused-electron-beam_2021}.} A quantitative model that consistently describes the internal composition of all FEBID NWs is not presented in this study, but \textcolor{\corrcol}{plausible} composition models for illustrative purposes are discussed in the supporting information \suppl{2}. In this analysis, the metallic atom content is measured as the average concentration of Co or Fe atoms in the centre band, corresponding to the volume indicated by dashed lines in Figure \ref{fig_model}, defined as the 20 nm wide area around the projected centre of mass in the EELS compositional map. To enable comparison, measurements of magnetic induction will also consider only the average magnetic field within this volume. 

\subsection{Measurement of NW elliptical cross-section shape}
The 3D sample geometry must be known to perform quantitative measurements of magnetic induction. Therefore, NW cross-sectional shapes are \textcolor{\corrcol}{approximated} by ellipses as shown in Figure \ref{fig_model}b, with the semi-major and semi-minor axes of the ellipse labelled $r_a$ \& $r_b$, respectively. They define a cylinder in 3D space, with coordinates $z = r_b \cos (p);~x = r_a \sin (p) $. Where $p$ is an arbitrary real-valued parameter and $z$ \& $x$ are coordinates. When the sample is tilted by an angle $\gamma$ during SEM imaging, it is equivalent to a rotation around the X'-axis of the SEM sample stage. Then, the observed NW width ($W$) is a tomographic projection \cite{toberg_quantitative_2020} of the ellipse, given by
\begin{equation}
    W(\gamma) = 2(r_a \cos(p_m) \sin(\gamma) + r_b \sin(p_m) \cos(\gamma)) .
    \label{eq_proj_width}
\end{equation}
Where $p_m = \arctan \left( \frac{r_b \cos(\gamma)}{r_a \sin(\gamma)} \right) $ is the ellipse parameter corresponding to the point farthest from the origin in the projection plane. The axes can be measured by imaging the sample at $\gamma=0\degree$ and $\gamma=90\degree$, providing values of  $W(0\degree) = 2 r_b$ and $W(90\degree) = 2 r_a$. When the model is accurate, $W$ measurements at intermediate $\gamma$ values are consistent with equation \ref{eq_proj_width}. In which case, the spatial variation of projected NW thickness during TEM imaging is given by \textcolor{\corrcol}{
\begin{equation}
    t_z (x) = 2 r_b \sqrt{1- \left( \frac{x}{r_a} \right) ^2}  .
    \label{eq_thickness}
\end{equation}}
This is needed to measure spatially resolved magnetic induction ($\boldsymbol{B}(x)$). $W$ measurements from both SEM and STEM were used to calculate $t_z$ because this approach reduces uncertainty and ensures consistency between instruments. The SEM images were acquired using an SE detector and have a 2.2~nm pixel size. STEM images were acquired with 2~nm pixel size using either an ADF detector or the EELS spectrometer. \textcolor{\corrcol}{For all NWs, the W(55°) measurements were within 5~nm of equation \ref{eq_proj_width}, which is equivalent to an $r_b$ measurement error lower than 2~nm and is assumed to be consistent with the model. Uncertainties in $r_a$ and $r_b$ are assumed equal to the pixel size and were propagated in calculations of $t_z$ and $\boldsymbol{B}$.}

\subsection{Magnetic characterisation}
Off-axis electron holography \cite{dunin-borkowski_electron_2019} is performed to reconstruct the magnetic phase shift ($\varphi_m$) of an electron wave passing through the sample and calculate $\boldsymbol{B}$. Electron holograms were acquired using an image aberration-corrected FEI Titan G2 60–300 TEM operated at 300 kV (Ernst Ruska Centre for Microscopy and Spectroscopy with Electrons) in magnetic field-free conditions (Lorentz mode).  Images and electron holograms were recorded using a direct electron counting Gatan K2 detector. A 100 V voltage was applied to the electron biprism, resulting in a hologram fringe spacing of approximately 3 nm. The total phase shift ($\varphi$) was reconstructed using Holoworks \cite{volkl_software_1995} and Holoview \cite{boureau_off-axis_2018} extensions for the Digital Micrograph Software package \cite{mitchell_scripting-customised_2005}. Maps of $\varphi$ are reconstructed with 5 nm spatial resolution, determined by the size of the virtual aperture selecting the hologram sideband. The phase maps showed 0.01 rad noise corresponding to the root-mean-square (RMS) difference between two vacuum reference images. 
Since the z-axis is defined as parallel to the TEM electron beam direction and the y-axis is parallel to the NW centreline, the Aharonov-Bohm effect \cite{aharonov_significance_1959} can be expressed as 
\begin{equation}
    \varphi=\ \varphi_{el}\left(V\right)+\ \varphi_m\left(A_z\right)=\ \frac{e}{\hbar v}\int V dz+\frac{e}{\hbar}\int A_zdz  .
\end{equation}
Where $\varphi_{el}$ and $\varphi_m$ are the electrostatic and magnetic contributions to the total phase shift ($\varphi$), respectively, $e$ is the charge of an electron, $\hbar$ is the reduced Planck constant, $v$ is the relativistic electron speed, $V$ is the electrostatic potential, $A_z$ is the z-component of the magnetic vector potential, and $\text{d}z$ is a path element along the z-axis. The integral is over the full trajectory of the beam. To separate $\varphi_m$ from $\varphi_{el}$, the NWs are magnetically saturated by tilting the sample to large opposite $\pm70\degree$ $\alpha$-tilt angles, applying an 1~T magnetic field using the objective lens of the microscope, and acquiring electron holograms at 0° $\alpha$-tilt angle for the opposite remnant states. Assuming that NW states after each saturation are equivalently opposite states of magnetisation ($\boldsymbol{M}$), the half-difference of the two $\varphi$ maps corresponds to $\varphi_m$ \cite{almeida_effect_2020}. \textcolor{\corrcol}{The $\boldsymbol{M}$ reversal is assumed to be complete because the $\varphi$ gradient in vacuum around the NWs had changed sign and kept the same magnitude.} Since $\boldsymbol{B} = \text{curl}(\boldsymbol{A})$, the thickness-averaged magnetic induction component parallel to the y-axis is given by 
\begin{equation}
    B_y(x)=\ -\frac{\hbar}{e\ t_z}\ \frac{\text{d}\varphi_m}{\text{d}x}   .
    \label{eq_By}
\end{equation}
where $t_z$ is the NW thickness in projection along the z-axis, given by equation \ref{eq_thickness}. The measurement of $B_y$ is used to compare the strength of magnetic induction fields created inside the NWs. 

\section{Results}

\subsection{Comparison of vertical NW composition}
\begin{figure}[ht]
  \centering
    \includegraphics[width=15cm]{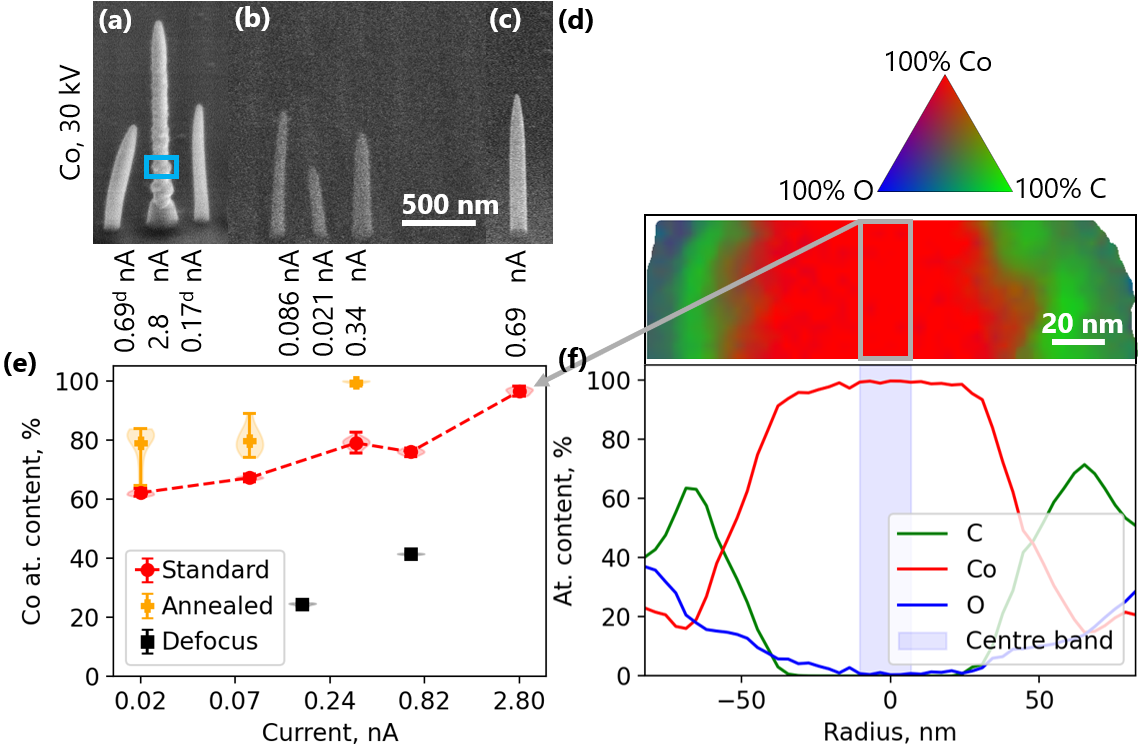}
  \caption{Comparison of vertical Co FEBID NWs deposited at 30~kV. (a-c) Vertical NWs from samples \vcot , \vcoanneal , and \acot, respectively. The scale is identical in all three images, annotations show the beam current during deposition, and the \textsuperscript{d} superscript identifies when the SEM beam was defocused. (d) EELS atomic composition map of the NW section corresponding to the blue box in (a). \textcolor{\corrcol}{(e) Atomic Co content (\%) in the NWs that are as deposited (standard), annealed, or deposited with a defocused beam. The error bars represent the range of Co content values measured in the centre band, and the violin plots represent the value histograms. The systematic uncertainty is 3\%. The abscissa scale is log$_2$. (f) Horizontal line-trace through (d), showing atomic content of all elements and the width of the centre band.}}
  \label{fig_curr}
\end{figure}

To ensure consistency of analysis for all deposits, the vertical FEBID NW samples \vcot$ $  and \vcoanneal$ $ (Figure~\ref{fig_curr}a-c) provided a compositional reference. Figure~\ref{fig_curr}d displays an EELS elemental map acquired from the blue square in Figure~\ref{fig_curr}a and reveals the metallic core and carbonaceous shell of the NW, as seen in previous works \cite{magen_focused-electron-beam_2021,pablo-navarro_purified_2018}. In Figure \ref{fig_curr}e, the thickness-averaged cobalt content measured from the central 20 nm band (Figure~\ref{fig_curr}d, rectangle) is compared to the other beam currents used for NWs in \vcot, \vcoanneal, and \acot$ $  samples. The systematic EELS fitting uncertainty is $<3\%$ for all points. In agreement with previous works, NWs with the highest metal content are deposited using higher beam currents and exhibit \textcolor{\corrcol}{uneven surface topography} \cite{fernandez-pacheco_magnetotransport_2009, muthukumar_spontaneous_2012, bertini_time-dependent_2006, pablo-navarro_purified_2018}. Figure \ref{fig_curr}e also demonstrates that annealing the \vcoanneal$ $ NWs at 350~°C for 30~min is an effective method of increasing mean \textcolor{\corrcol}{Co content}, and correctly focusing the beam is important for depositing with high initial Co content \cite{botman_creating_2009, puydinger_dos_santos_annealing-based_2016}. \textcolor{\corrcol}{Supporting Information \suppl{1} contains composition maps of all NWs, which are used for comparison in this subsection. It is assumed that the right side of the carbon-rich shell in Figure~\ref{fig_curr}d is thicker primarily due to surface topography, as EELS maps acquired at a higher elevation along the NW indicate similar composition with reduced asymmetry. Furthermore, composition maps for vertical NWs deposited left-to-right with spacings from 300~nm to 500~nm do not show consistently increased shell thickness on the right side that might indicate codeposition induced by proximity effects \cite{burbridge_proximity_2009}.} In summary, the compositional analysis agrees with previous observations that the deposit metal content scales generally with electron beam current \cite{de_teresa_review_2016}, but high currents result in less consistent surface topologies. 

\subsection{Calibration of oblique NW shape and composition measurements}
\begin{figure}[h]
  \centering
    \includegraphics[width=15cm]{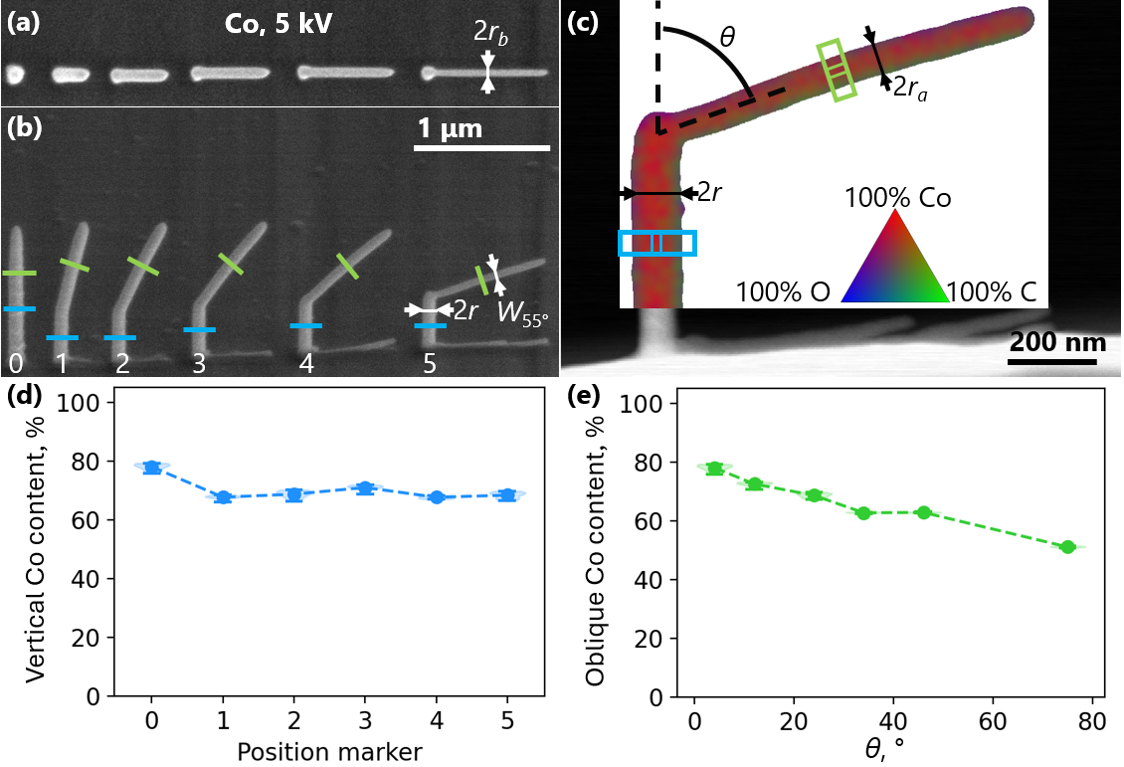}
  \caption{Measurement of cross-section shape and Co content for the \acof$ $ sample. (a) 0° and (b) 55° tilt SEM images of NWs, respectively. The lines indicate the locations of high-resolution STEM EELS scans. (c) STEM EELS map displayed over a STEM ADF map of NW No.~5. The boxes indicate the locations of high-resolution STEM EELS scans, $\theta$ is the NW growth angle, and black lines correspond to width measurements relating to equation \ref{eq_proj_width}. (d)  Co content in NW regions corresponding to the horizontal blue lines in (b), used to verify consistency. (e)  Co content in NW regions corresponding to the angled green lines in (b).}
  \label{fig_one_ang}
\end{figure}

To enable comparison, all NWs are characterised using the same methods. Since NW cross-sections can be elliptical, the NW diameter is measured twice by tilting the SEM stage and acquiring images at $\gamma = 0\degree$ and $\gamma = 55\degree$ tilt, shown respectively in Figures \ref{fig_one_ang}a and \ref{fig_one_ang}b. The projected NW width is measured as the full-width at half-maximum (FWHM) of the intensity line profiles in the SEM images, as shown by the lines corresponding to $W(55\degree)$ and $2 r_b$. 

STEM EELS elemental maps are acquired with 2~nm resolution to characterise areas of interest, as shown in Figure \ref{fig_one_ang}c. The sample orientation during STEM imaging is equivalent to a $\gamma = 90\degree$ tilt, therefore the projected NW widths in the EELS maps can be used to determine $r_a$. Additionally, the radius of the calibration element ($r$) is the same in all projections, thus $r$ measurements can be compared between SEM and STEM EELS. Based on EELS thickness measurement, the SEM detects the NW edge at the position where the projected material thickness is $\sim0.5$~nm, which is significantly below the pixel size and does not require correction. Furthermore, the $W(55\degree)$ measurement is within two pixel-widths (5 nm) of equation \ref{eq_proj_width}, showing that the NW cross-sectional shape is consistent with an ellipse that has semi-axes $r_a$ and $r_b$. Comparisons between NWs will be presented in a later section.

The deposition consistency is evaluated by ensuring all nanostructures include a NW calibration element deposited with a stationary beam, such as the vertical sections in Figure~\ref{fig_one_ang}b. EELS maps are used to compare Co content in the calibration element, as shown in Figure~\ref{fig_one_ang}d. The histograms of Co content measurements in NW centre bands, corresponding to the rectangular boxes in Figure~\ref{fig_one_ang}c, are shown by shaded areas between error bars (violin plots). The $\theta=0\degree$ NW is the outlier as the SEM beam is stationary, thus it transmits through the NW and irradiates the bottom layers for the entire deposition time. Consistent calibration sections indicate that deposition parameters do not vary significantly across nanostructures, enabling comparison of the oblique NW sections. The growth angle $\theta$ is defined relative to the SEM optical axis, as shown in Figure~\ref{fig_one_ang}c. Co content in the oblique NWs is compared in Figure~\ref{fig_one_ang}e. This process is repeated to find correlations between metal content and $\theta$ in all oblique NW samples.

\subsection{Correlation between composition and growth angle}
\begin{figure}[h]
  \centering
    \includegraphics[width=15cm]{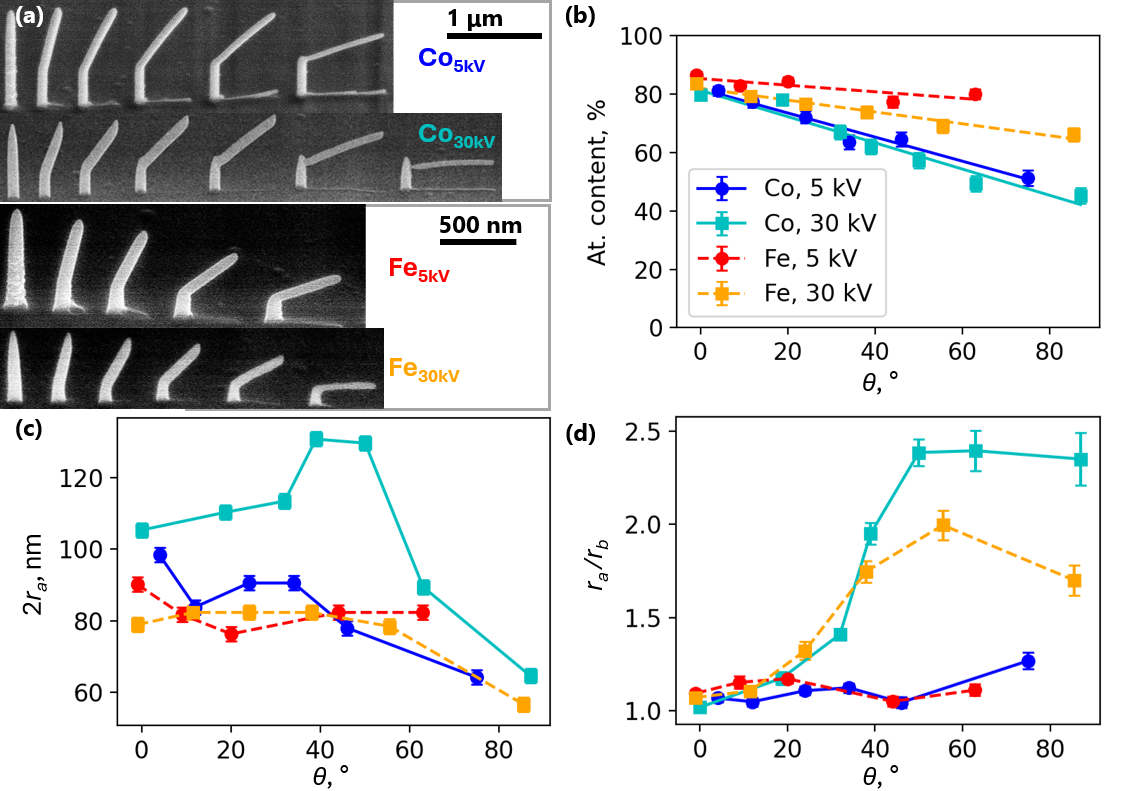}
  \caption{Variation in oblique NW composition and shape as a function of growth angle. (a) SE SEM images of the \acof, \acot, \afef, and \afef$ $ NWs. (b) Mean atomic content of metal (Co or Fe) in the NW centre bands as a function of $\theta$. Error bars in (b) are not drawn to improve graph readability, but would represent 3\% systematic uncertainty and less than 2\% random error for each point. \textcolor{\corrcol}{The lines represent linear regression fits for each dataset.} (c) The length of the major axis of the elliptical NW cross-section, as a function of $\theta$. Error bars correspond to STEM image pixel size. (d) Ratio of the major and minor axes of the cross-section, as a function of $\theta$. Error bars correspond to one standard deviation of random error. For all tested precursor gases, accelerating voltages, and beam currents, the NW metal content and cross-sectional area decrease with increasing beam movement speed.}
  \label{fig_ang}
\end{figure}

Having established fabrication and characterisation methods consistent with the expected results for vertical NWs, oblique NWs are characterised. Figure \ref{fig_ang} presents the variation of NW composition and shape as a function of $\theta$, for samples \acof, \acot, \afef, and \afet$ $ (Figure~\ref{fig_ang}a), which were fabricated by \textcolor{\corrcol}{laterally moving} the SEM beam during deposition. Figure \ref{fig_ang}b shows that the mean metal at. content (Co or Fe) in the 20 nm centre band declines with increasing $\theta$ for all NWs. \textcolor{\corrcol}{The average metal content loss per degree of growth angle is represented by the slope of a linear regression fit. It is $(0.11 \pm 0.05)$\%/° for the \afef$ $ sample, $(0.20 \pm 0.02)$\%/° for \afet, $(0.41 \pm 0.04)$\%/° for \acof, and $(0.45 \pm 0.05)$\%/° for \acot.} The length of $r_a$ is shown in Figure \ref{fig_ang}c, revealing that the NW cross-sectional area shrinks during \textcolor{\corrcol}{lateral} beam movement. Figure \ref{fig_ang}d displays the $r_a/r_b$ ratio as a function of $\theta$, indicating that the NW cross-section becomes more elliptical if a 30 kV accelerating voltage is used and the beam is \textcolor{\corrcol}{moved laterally}, confirming previous works \cite{winkler_shape_2021}.  \textcolor{\corrcol}{Comparison between $r_a$ and $r_b$ is shown in Supporting Information Figure \suppl{S11}, indicating that the minor axis also decreases as a function of $\theta$.} The results show that \afef$ $ maintains the highest metal content and the most circular cross-section shape when $\theta$ is increased. \textcolor{\corrcol}{Additionally, Supporting Information Figure \suppl{S12} shows a correlation between at. content measurement and NW cross-sectional area. Overall, \afef$ $ is the only dataset where $\theta$, $r_a$, and metal at. content measurements are weakly correlated.}

\subsection{Micromagnetic simulations of cylindrical FEBID NWs}
The micromagnetic simulations were performed using Mumax3 \cite{vansteenkiste_design_2014, joos_tutorial_2023} to determine $\boldsymbol{M}$ configurations that are energetically stable in cylindrical FEBID NWs comprising nanocrystalline \textcolor{\corrcol}{or amorphous} impure Co. To qualitatively evaluate possible $\boldsymbol{M}$ configurations, FEBID NWs can be approximated as uniform cylinders \cite{fullerton_design_2025, donnelly_complex_2022}. The exchange stiffness is $A_{ex} = 1.5 \cdot 10^{-11}$ J/m and the saturation magnetisation is $M_s = 8 \cdot 10^5$ A/m, based on previous FEBID simulations \cite{fullerton_design_2025} and measurements of Co samples of different purities \cite{pablo-navarro_purified_2018, silinga_model-based_2025, donnelly_complex_2022}. The simulation parameters for nanocrystalline impure Fe are similar \cite{rodriguez_influence_2015} and produce equivalent results. A cubic mesh with 2.5 nm voxel size is used for an exchange length $l_{ex}=\ \sqrt{\frac{2A_{ex}}{\mu_0{M_s}^2}}\approx6 $ nm. 

To investigate magnetic configurations in FEBID NWs, 1 \textmu m long cylinders of  Co and Fe are simulated using micromagnetics. The diameters are varied in the range from 60 nm to 240 nm, and the initial $\boldsymbol{M}$ states are aligned parallel to the y-axis \textcolor{\corrcol}{(magenta arrow in Figure~\ref{fig_febid})}, such that the simulation emulates a relaxation after being subject to a saturating external field \cite{joos_tutorial_2023}. Corresponding energetically stable configurations are shown in \textcolor{\corrcol}{Supporting Information Figure~\suppl{S13}}. Whilst magnetic flux closure is always visible at the cylinder ends, the majority of the volume is magnetised parallel to the long axis if the cylinder diameter is below 160 nm. For diameters above 160 nm, various vortex configurations are observed. \textcolor{\corrcol}{Micromagnetic simulations of NWs with elongated cross-sections corresponding to geometries in the experimental dataset show that oblique NWs are also predominantly magnetised parallel to the long axis. The $\boldsymbol{M}$ configuration corresponding to the most elliptical NW with $r_a = 62.5$~nm and $r_b = 25$~nm is shown in the Supporting Information Figure \suppl{S14}}. Based on these simulations, the oblique NWs in Figure \ref{fig_ang} are expected to be uniformly magnetised. Hence, simulations show that the thickness-average of $B_y$ calculated from $\varphi_m$ using equation \ref{eq_By} corresponds to the average magnitude of $\boldsymbol{B}$.

The recorded $\overline{B_y}$ values represent a measurement of average $\boldsymbol{B}$ magnitude in volumes where the demagnetising field is non-negligible, and the NW material comprises both ferromagnetic and paramagnetic regions. However, since the NWs are unidirectionally magnetised and have a length-to-projected-diameter ratio greater than 5:1, $\overline{B_y}$ is indicative of the saturation induction $B_0$, which is an intrinsic material property. Based on modelling of the demagnetising fields, the average $B_0$ is expected to be from 10\% to 20\% greater than $\overline{B_y}$ for such NW geometries \cite{akhtari-zavareh_magnetic_2017}, but would require 3D characterisation to identify variations in non-ferromagnetic NW shell thickness that limit the accuracy of micromagnetic simulations.

\subsection{Magnetic Induction in oblique NWs}
\begin{figure}[p]
  \centering
    \includegraphics[width=13cm]{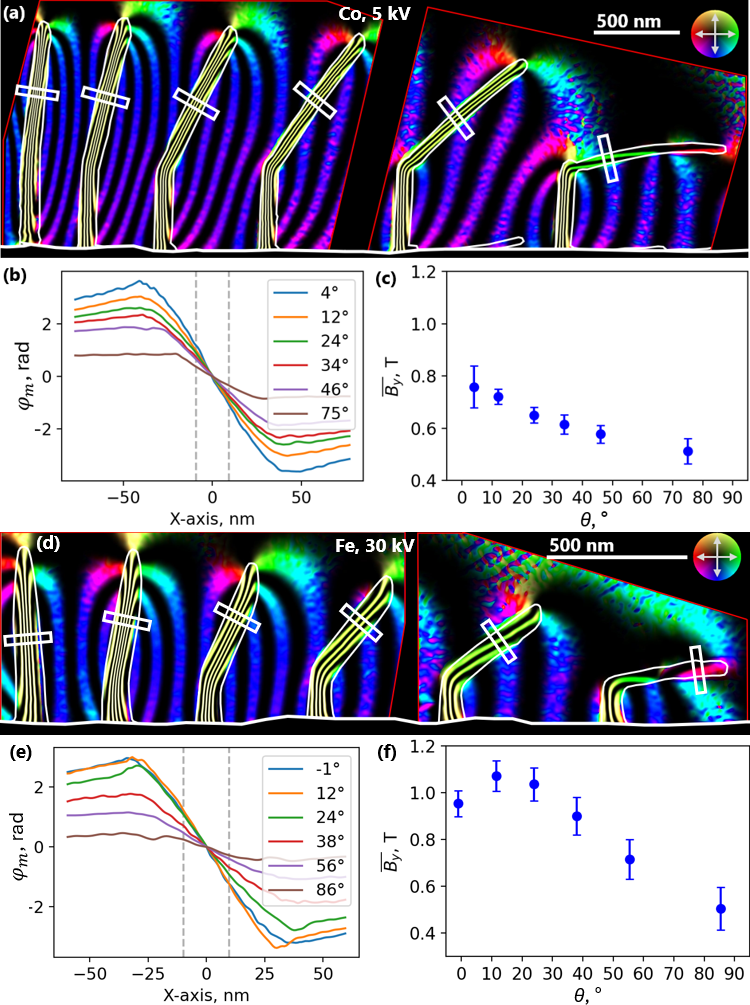}
  \caption{\textcolor{\corrcol}{Magnetic characterisation of the \acof$ $ (a-c) and \afet$ $ (d-f) samples. (a), (d) Magnetic induction maps with contour spacings of $2\pi/5$~rad. (b), (e) Line traces of $\varphi_m$ are from areas corresponding to the white boxes in (a) and (d), respectively. The measured $\varphi_m$ gradient is consistently smaller for higher $\theta$. The centre bands are denoted by dashed lines. (c), (f) Correlations between $\theta$ and $B_y$ in the centre bands.}}
  \label{fig_co_holo}\label{fig_fe_holo}
\end{figure}

Figure \ref{fig_co_holo} presents the magnetic imaging of NWs in  \acof$ $ and \afet$ $ samples as a function of growth angle $\theta$, and by implication metal content, respectively. These samples consist of nanowires in which the metal content shows a similar dependence on $\theta$, even though during deposition the accelerating voltage, beam current, and precursor gas were different for each sample. Figure \ref{fig_co_holo}a shows magnetic induction maps reconstructed from electron holography of the \acof$ $ NWs deposited using a 340~pA beam current. To create magnetic induction maps, the $\varphi_m$ images underwent Gaussian smoothing and the cosine was amplified (x5) to produce magnetic phase contours with $2\pi/5$~rad spacing, and colour wheels are used to show the projected $\boldsymbol{B}$ direction. Figure \ref{fig_co_holo}b shows line traces of $\varphi_m$ as a function of $\theta$, from which, after adjusting for variations in NW cross-sectional area, the mean $B_y$ value in the 20 nm centre bands was calculated (Figure \ref{fig_co_holo}c). The error in $\overline{B_y}$ represents the combined uncertainty of projected NW thickness measurements and the standard deviation of $B_y$ in the centre bands. It should be noted that the lowest-purity NW deposited at $\theta > 70\degree$ was observed to warp during STEM EELS mapping, creating a site where a magnetic domain wall regularly forms after saturating the sample. Hence, the $\varphi_m$ line trace was acquired off-centre, so it is in the middle of a uniformly magnetised region.

For comparison, Figures \ref{fig_fe_holo}d-f show magnetic induction maps of the \afet$ $ NWs deposited using a beam current of 2800~pA (Figure \ref{fig_fe_holo}d), line traces of $\varphi_m$ as a function of $\theta$ (Figure \ref{fig_fe_holo}e) and the calculated mean $B_y$ value in the centre bands (Figure \ref{fig_fe_holo}f). In both \acof$ $ and \afet$ $ samples, the NWs grown at larger angles are observed to contain a lower percentage of metal atoms, and the $\overline{B_y}$ value reduction is 2 mT and 7 mT per degree of $\theta$, respectively. As noted above, the outliers to this trend occur at $\theta \approx 0\degree$ in both datasets due to the stationary electron beam transmitting through the length of the NW for the entire duration of deposition, thereby altering the geometry and composition of the lower layers.

\begin{figure}[ht]
  \centering
    \includegraphics[width=15cm]{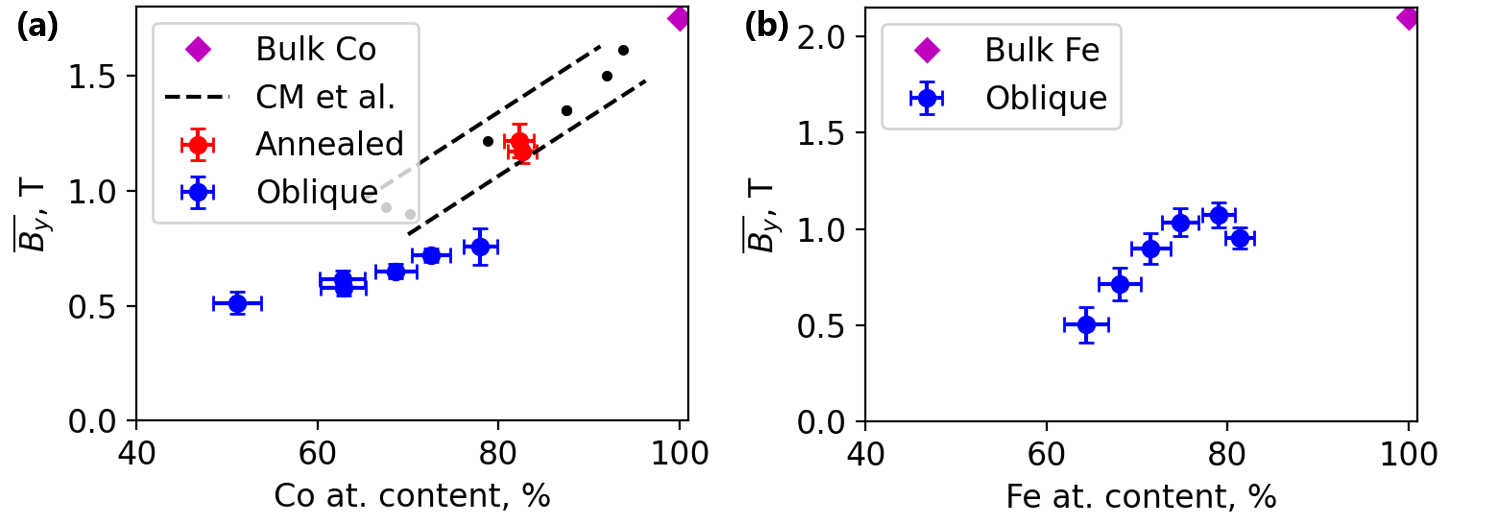}
  \caption{Correlations between NW metal content and associated magnetic properties. (a) Correlation between mean $B_y$ and atomic Co content for FEBID NWs. \textcolor{\corrcol}{As-deposited oblique NWs are shown by blue circles, annealed vertical NWs are shown by red circles, and bulk material reference values are shown by magenta diamonds.} Black dots and dashed lines indicate the data and range of uncertainties in previous works \cite{magen_focused-electron-beam_2021}, respectively. (b) Correlation between mean $B_y$ and Fe content for oblique FEBID NWs. There is a positive correlation between metal content and the strength of magnetic induction created by NWs. Comparison with previous works shows that the correlation is affected by structural differences within the NWs, resulting from thermal annealing \textcolor{\corrcol}{and lateral beam movement. Adapted from \cite{magen_focused-electron-beam_2021}, available under a CC-BY 4.0 license, copyright 2021 C. Magén, J. Pablo-Navarro, and J. M. De Teresa.}}
  \label{fig_correl}
\end{figure}

Figure \ref{fig_correl} summarises the correlation between NW metal content and $\overline{B_y}$ measurement. The data in Figures \ref{fig_co_holo}c and \ref{fig_fe_holo}f are plotted with respect to the metal content in each oblique NW and are compared with the literature. The two red data points in the top right of Figure~\ref{fig_correl}a correspond to the thermally annealed NWs and agree with previous works \cite{magen_focused-electron-beam_2021, pablo-navarro_purified_2018} characterising annealed Co FEBID NWs (Figure \ref{fig_correl}a, black dots and dashed lines). The underlying magnetic induction maps for annealed NWs are available in the Supporting Information Figure~\suppl{S15}. All other blue data points correspond to \acof$ $ NWs, which are not annealed, and lie outside the range of the referenced study. For FEBID NWs, $\overline{B_y}$ is smaller than 1.75 T $B_0$ for pure bulk Co \cite{snoeck_quantitative_2003, serrano-ramon_ultrasmall_2011}. For the \afet$ $ NWs, the values of $\overline{B_y}$ in Figure \ref{fig_correl}b are lower than the 2.1 T value measured for pure bulk iron \cite{crangle_magnetization_1971}. $\overline{B_y}$ is lower than $B_0$ because it is affected by the demagnetising field and represents the thickness-average in NWs with radially varying composition, but it is indicative of the variation in effective $\boldsymbol{B}$, which scales with the percentage of metal atoms in the material. Overall, the $\boldsymbol{B}$ field created in ferromagnetic NWs is affected by growth angle due to changes in elemental composition.

\section{Discussion}
In this work we have fabricated and analysed oblique FEBID NWs to determine improved parameters for depositing ferromagnetic material in oblique geometries. Characteristic FEBID NW variations in response to changes in deposition parameters were observed, including increases in metal content with increased electron beam current, precise focussing of the electron beam, or thermal annealing of the NWs at 350~\textdegree C for 30 min \cite{pablo-navarro_purified_2018, randolph_focused_2006, cordoba_high-purity_2016}. These conditions were all considered to find optimal beam currents for fabricating obliquely angled ($0\degree \leq \theta \leq 90\degree$) Co and Fe NWs at 5~kV and 30~kV accelerating voltages. In all samples, increasing the electron beam \textcolor{\corrcol}{lateral movement} speed characteristically increases $\theta$, but reduces NW cross-section area, metal content, and the magnitude of $\boldsymbol{B}$. \textcolor{\corrcol}{Oblique NWs that were the most symmetrical and had the highest metal content} (without annealing) were deposited using the combination of the lowest tested accelerating voltage (5 kV), the highest beam current (2.8 nA), and the Fe$_2$(CO)$_9$ precursor. 

The beam current is increased by increasing the spot-size \cite{williams_lenses_2009}, which enables more electrons to pass through the condenser aperture at the expense of beam broadening. The corresponding elevation in precursor ionisation and beam-induced heating promotes deposition with higher atomic metal content, but NW diameter is increased. The highest beam current of 2.8 nA used in this study resulted in a NW diameter lower than 150 nm. Micromagnetic simulations demonstrated that, in 1 \textmu m long Co or Fe cylinders with diameters $\leq 150$~nm, $\boldsymbol{M}$ is mostly uniaxially oriented, because small diameters strengthen magnetic shape anisotropy. Hence, small-diameter NWs are investigated due to the enhanced magnetic signal associated with uniaxial magnetic configurations, which is advantageous for spintronic device applications \cite{parkin_magnetic_2008}. Additionally, excessively high beam currents can lead to unpredictable \textcolor{\corrcol}{NW surface topographies} (Figure \ref{fig_curr}a). Especially for the Fe$_2$(CO)$_9$ precursor, which dissociates to Fe(CO)$_5$ \cite{bertini_time-dependent_2006} during irradiation, FEBID can produce randomly oriented crystalline iron composed of single-crystal grains up to 1 \textmu m in length \cite{hochleitner_electron_2008}. Consequently, the maximum viable beam current is limited by the NW diameter and required growth predictability.

The optimal electron beam acceleration voltage is often sample-specific because electron energy affects the optical power of  SEM lenses, and alters the scattering interaction cross-sections in the NW and the substrate \cite{williams_lenses_2009, castillo-rico_stopping_2021}. In this study, a minimum accelerating voltage of 5 kV was used, as lower voltages compromised imaging resolution and reduced the ability to achieve good focus. At 5~kV, the cross-sections of oblique NWs are circular (Figure \ref{fig_ang}d) since the interaction volume is comparable to the NW diameter and secondary electrons are generated in a localised spot \cite{joy_low_1996}. At 30~kV, highly elliptical cross-sections are observed due to secondary electrons being generated along a line as the beam transmits through the NW. Additionally, the vertical growth rate in oblique NWs is larger at 5~kV compared to 30~kV (Figure \ref{fig_ang}a), whereas Table \ref{table1} shows an inverse relationship of vertical NWs growing faster at 30~kV than at 5~kV, when using high beam currents. This is attributed to the smaller interaction cross-sections of 30~kV electrons \cite{ashley_calculations_1976}, resulting in electrons imparting less energy during transmission through the sample surface. Despite sample specificity, the 0.1 – 5 kV range is considered advantageous for FEBID of oblique NWs due to the smaller interaction volume, where the incident beam provides an approximately point-like source of secondary electrons \cite{cordoba_high-purity_2016, castillo-rico_stopping_2021} and leads to more consistent deposition.  

Of the tested precursors, the Fe$_2$(CO)$_9$ molecule is more stable than Co$_2$(CO)$_8$, as indicated by the higher required beam currents (Table \ref{table1}) and threshold temperature for TD \cite{connor_high_1973}. In Figure \ref{fig_ang}b, the elemental composition of Fe NWs, in comparison to Co, is less affected by variations in the electron beam \textcolor{\corrcol}{movement} speed. This is attributed to improved confinement of deposition reactions, as molecules outside the immediate vicinity of the beam are less impacted by scattered electrons or conducted heat. It should be noted that other FEBID precursor gases are available \cite{de_teresa_review_2016}, and HCo$_3$Fe(CO)$_{12}$ has been used to successfully deposit oblique ferromagnetic NW structures \cite{keller_direct-write_2018}.

NW arrays \cite{fullerton_design_2025}, artificial spin ices \cite{keller_direct-write_2018}, and 3D data storage \cite{gu_three-dimensional_2022} prototypes can be fabricated with FEBID. Achieving uniform composition throughout these structures is essential to ensure reproducible and reliable device performance. We have demonstrated that $>75\%$ metal content can be maintained during deposition of oblique NWs angled $0\degree<\theta<60\degree$, when the deposition parameters are chosen to confine the reaction volume to dimensions no larger than the NW diameter, whilst using the highest viable beam current. However, NWs grown at angles $\theta > 60$° from the optic axis exhibit a twofold decrease in magnetic signal ($\nabla \varphi_{m}$) due to reduced NW cross-sectional area. It may be possible to correct this effect by increasing the NW diameter in associated CAD modelling software \cite{skoric_layer-by-layer_2020}. The lowest observed decrease in metal at. content was, on average, $(0.11 \pm 0.05)$\% per degree of $\theta$. Better material uniformity may be achieved by investigating how precursor gas concentration affects oblique NWs, or by using accelerating voltages below 5 kV, which would require the introduction of additional sample features to enable fine focus. \textcolor{\corrcol}{Alternatively, the substrate may be tilted to deposit oblique nanostructures with less lateral beam movement.} Additionally, the metal content and resultant magnetic induction (Figure \ref{fig_correl}) may be increased by thermal annealing observed in vertical NWs (Figure \ref{fig_curr}), but the optimal temperature for preserving nanostructure uniformity would need to be determined. \textcolor{\corrcol}{Figure \ref{fig_correl}a has shown that metal atom content measured in projection is not the only parameter affecting the magnetic induction created by NWs, and this may be investigated by characterising magnetic fields for a wider range of FEBID nanostructures.} Overall, selecting deposition parameters to limit the reaction volume and use the highest viable electron beam current \textcolor{\corrcol}{can} improve the compositional uniformity of oblique ferromagnetic FEBID NWs, which is beneficial for the fabrication of uniform 3D nanostructures.

\section{Conclusions}
Strategies are discussed for optimising deposition parameters to improve material uniformity in 3D ferromagnetic FEBID nanostructures. This work demonstrates inherent variation in the material properties of oblique FEBID NWs deposited with a \textcolor{\corrcol}{laterally moving} electron beam, as compared to a stationary beam, and shows that these properties can be tuned by adjusting the deposition parameters. Higher-energy electron beams penetrate further into a material, which can delocalise the deposition reactions by irradiating a larger volume. FEBID of oblique Co NWs using a 30 kV beam reveals a reduction in cross-section area and a decrease in metal content by up to 0.4\% per degree of growth angle relative to the optic axis. TEM characterisation confirms that the reduction of NW purity results in a corresponding weakening of the magnetic induction. Ensuring uniform NW composition enhances the consistency of \textcolor{\corrcol}{functional 3D nanostructures} fabricated with FEBID. Consequently, oblique NWs fabricated using different combinations of SEM's electron beam current, accelerating voltage, and precursor gas type were characterised to assess compositional variations. By using the \textcolor{\corrcol}{deposition} conditions of: 5~kV accelerating voltage; 2.8~nA beam current; and the Fe$_2$(CO)$_9$ precursor gas, the average decrease in NW metal content was limited to \textcolor{\corrcol}{$(0.11 \pm 0.05)$\%/°}, and the NW cross-sectional shape was preserved for growth angles up to 60°. In summary, it was shown that metal content in oblique NWs is dependent on the growth angle, but variation in material properties can be minimised by using a precursor gas and accelerating voltage that confine the deposition reactions to an interaction volume no larger than the NW diameter.

\section*{Acknowledgements}

The authors acknowledge funding from the Engineering and Physical Science Research Council (Grant No. EP/W524359/1, EP/X025632/1), the European Research Council under the European Union's Horizon 2020 Research and Innovation Programme (Grant No. 856538, project '3D MAGiC'), the Deutsche Forschungsgemeinschaft (Project-ID 405553726 TRR270), and the Helmholtz fellowship. Aurys Šilinga thanks Fred Rendell-Bhatti and Wyn Williams for constructive discussions.

\section*{Data availability}
The data underlying this study are openly available in Enlighten at \url{http://dx.doi.org/10.5525/gla.researchdata.2427}.

\section*{Supporting information}
\label{sec_supporting_information}
The following files are available free of charge.
\begin{itemize}
  \item Supporting\_information.pdf: EELS scans underlying the datapoints in Figure \ref{fig_ang}, a discussion about generating the model in Figure \ref{fig_model}, comparison of high-loss EELS spectra from NWs of various projected thicknesses, additional plots showing correlations between NW geometry and atomic composition, micromagnetic simulations of NWs, and magnetic induction maps of vertical annealed NWs underlying calibration datapoints in Figure \ref{fig_correl}a.
\end{itemize}

\printbibliography

\end{document}